\documentclass{article}
\usepackage{ijcai24}

\usepackage{times}
\usepackage{soul}
\usepackage{url}
\usepackage[hidelinks]{hyperref}
\usepackage[utf8]{inputenc}
\usepackage[small]{caption}
\usepackage{graphicx}
\usepackage{amsmath}
\usepackage{amsthm}
\usepackage{booktabs}
\usepackage{algorithm}
\usepackage{algorithmic}
\usepackage[switch]{lineno}
\usepackage[english]{babel}

\theoremstyle{definition}
\newtheorem{definition}{Definition}[section]

\title{Tell me more: Intent Fulfilment Framework for Enhancing User Experiences in Conversational XAI}

\author{
Anjana Wijekoon$^1$
\and
David Corsar$^1$
\and
Nirmalie Wiratunga$^1$
\and
Kyle Martin$^1$
\and
Pedram Salimi$^1$
\affiliations
$^1$School of Computing, Robert Gordon University\\
\emails
\{a.wijekoon1, d.corsar1, n.wiratunga, k.martin3, p.salimi\}@rgu.ac.uk
}

\begin{document}

\maketitle

\begin{abstract}
The evolution of Explainable Artificial Intelligence~(XAI) has emphasised the significance of meeting diverse user needs.
The approaches to identifying and addressing these needs must also advance, recognising that \textit{explanation experiences} are subjective, user-centred processes that interact with users towards a better understanding of AI decision-making. 
This paper delves into the interrelations in multi-faceted XAI and examines how different types of explanations collaboratively meet users' XAI needs.
We introduce the Intent Fulfilment Framework~(IFF) for creating explanation experiences. The novelty of this paper lies in recognising the importance of ``follow-up'' on explanations for obtaining clarity, verification and/or substitution. 
Moreover, the Explanation Experience Dialogue Model integrates the IFF and ``Explanation Followups'' to provide users with a conversational interface for exploring their explanation needs, thereby creating explanation experiences.
Quantitative and qualitative findings from our comparative user study demonstrate the impact of the IFF in improving user engagement, the utility of the AI system and the overall user experience.
Overall, we reinforce the principle that ``one explanation does not fit all'' to create explanation experiences that guide the complex interaction through conversation.

\end{abstract}

\section{Introduction}

Artificial Intelligence (AI) is being increasingly integrated into all aspects of society. As such, there is a growing armoury of explanation techniques to provide transparency and justifications for AI's underlying decision-making processes. 
Recent work has shown that conversational interactions are effective at understanding the user's explanation needs~(i.e. intents) and utilising explanation techniques to address them~\cite{madumal2019grounded,malandri2023convxai}. 
However, the current state-of-the-art overlooks the dynamics between receiving explanations and the essential follow-up interactions, which are prompted by changes in the user's mental model that can improve understanding, build trust, and significantly improve overall user experience.

Consider a loan applicant with a rejected loan application exploring their next steps with an Explainable AI~(XAI) system. 
A feature attribution explanation will show which details of their application attributed the most to a rejected outcome. The applicant may follow up by requesting additional information to \textit{complement} their understanding of the attributions; or to \textit{validate} the explanation by requesting feature attribution explanation from another algorithm; or an alternative explanation, such as one involving comparisons with similar previous loan applications. Such interactions necessitate a model that enables structured interactions, precisely understanding and addressing the user's specific follow-up needs to enhance engagement.

In this paper, we delve into the interrelations between explanation intents and explanations to introduce the Intent Fulfilment Framework~(IFF) as the general set of guidelines for creating explanation experiences. 
Specifically, we ask the research question: \textit{``How to model relationships between explanation intents and explanation types to create conversational explanation experiences?''}.
Addressing this, we make the following contributions:
\begin{itemize}
\item Intent Fulfilment Framework~(IFF): an ontology-guided typological classification of explanation intents, user questions, explanation types and relations between explanations;
\item Explanation Experience Dialogue Model~(EEDM) that implements IFF and Followup dialogue constructs to create conversational explanation experiences; and 
\item the findings of a comparative user study that evaluated the impact of the IFF on user engagement, utility of the AI system and overall user experience.
\end{itemize}

The rest of the paper is organised as follows. Section~\ref{sec:related} presents the related work. In Section~\ref{sec:onto} we present the IFF, followed by the formalisation of the EEDM in Section~\ref{sec:adf}. Section~\ref{sec:study} presents the details and findings of the comparative user study and discusses broader implications. Finally, we offer conclusions in Section~\ref{sec:conc}.

\section{Related Work}
\label{sec:related}
\subsection{Multi-faceted XAI}
\label{sec:related-xai}
XAI is a mature field of study where the goal of explanations is to provide insights into how a model works, why a certain decision was made or what generally impacts model decisions~\cite{arrieta2020explainable}. 
Explanations should cater to different types of users, and their explanation needs~\cite{vilone2021notions,liao2020questioning}. 
Increasingly it is realised that XAI is not one-shot, and should be interactive, and able to provide multiple explanations considering diverse user intents~\cite{sokol2020one,liao2020questioning}. 

In the literature, we find several cases that explored the relationships between intents and explanation techniques. 
Some systems are driven by domain-specific knowledge, tailored for particular use cases within fields such as medicine~\cite{schoonderwoerd2021human} or finance~\cite{sokol2020one}, incorporating insights from experts in the AI system's area of application.
More commonly, researchers use a data-driven approach to compare different types of explanations with target groups to make recommendations~\cite{tsai2021exploring,sivaraman2023ignore,kim2023help,baumann2021choices,panigutti2022understanding,anik2021data}. 
While the relationships between explanations and intents are use-case-specific, we argue that there is a significant gap due to the absence of a general set of guidelines. 
Such a general framework will reduce the entry barrier for AI system owners and the cost of curating a use-case-specific ``explanation strategy''. 
The taxonomy proposed in~\cite{arya2019one} is an attempt at bridging this gap, but it lacks user-centred design and evaluation and therefore fails to capture the influence of diverse explanation types and relationships between them. 
In this paper, we address this gap by presenting a typology classification between intents, user questions and explanation types. 

\subsection{Conversational XAI}
\label{sec:related-conv}
Conversation is an upcoming medium for implementing interactive XAI, offering an alternative to graphical or text-based user interfaces~\cite{chromik2021human,madumal2019grounded,sokol2020one,malandri2023convxai}.
The first known formalisation of dialogical interactions for XAI was the Explanation Dialogue Model~\cite{madumal2019grounded} where their data-driven approach recognised the user's ability to engage in an argument following an explanation. ConvXAI~\cite{malandri2023convxai} extended this model to include ``clarification'' dialogue type that provides additional information to enhance the usefulness of the explanation.
Previous work lacks specificity regarding the type of additional information~\cite{madumal2019grounded} or clarification~\cite{malandri2023convxai} leading to implementations reliant on expert knowledge. We address this oversight by recognising various follow-up needs and utilising alternative explanation techniques to meet them. 
Accordingly, we introduce 3 ``follow-up'' types that provide information or explanations to complement, replace and/or validate an explanation that is generated from explanation techniques or by leveraging the knowledge of the AI system development. The specificity of these Followup types promotes the reuse of existing techniques and diminishes the excessive need for external expert knowledge. 
Following previous literature we formalise the proposed dialogue constructs using the Agent Dialogue Framework~\cite{mcburney2002games}.
\section{Intent Fulfilment}
\label{sec:onto}
This section explores the concepts and their relationships related to intent fulfilment in XAI to present the Intent Fulfilment Framework~(IFF). 

\subsection{Ontology}

\begin{figure}[!t]
\centering
\includegraphics[width=.48\textwidth]{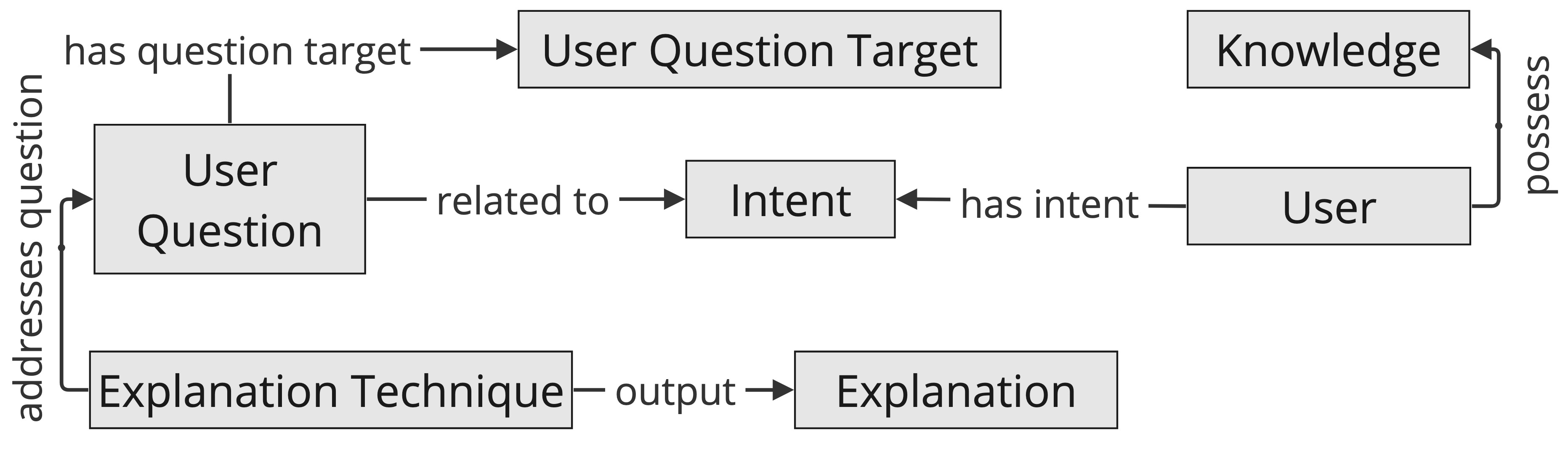}
\caption{Intent Fulfilment Ontology}
\label{fig:onto}
\end{figure}

\begin{figure*}[!t]
    \centering
    \includegraphics[width=\textwidth]{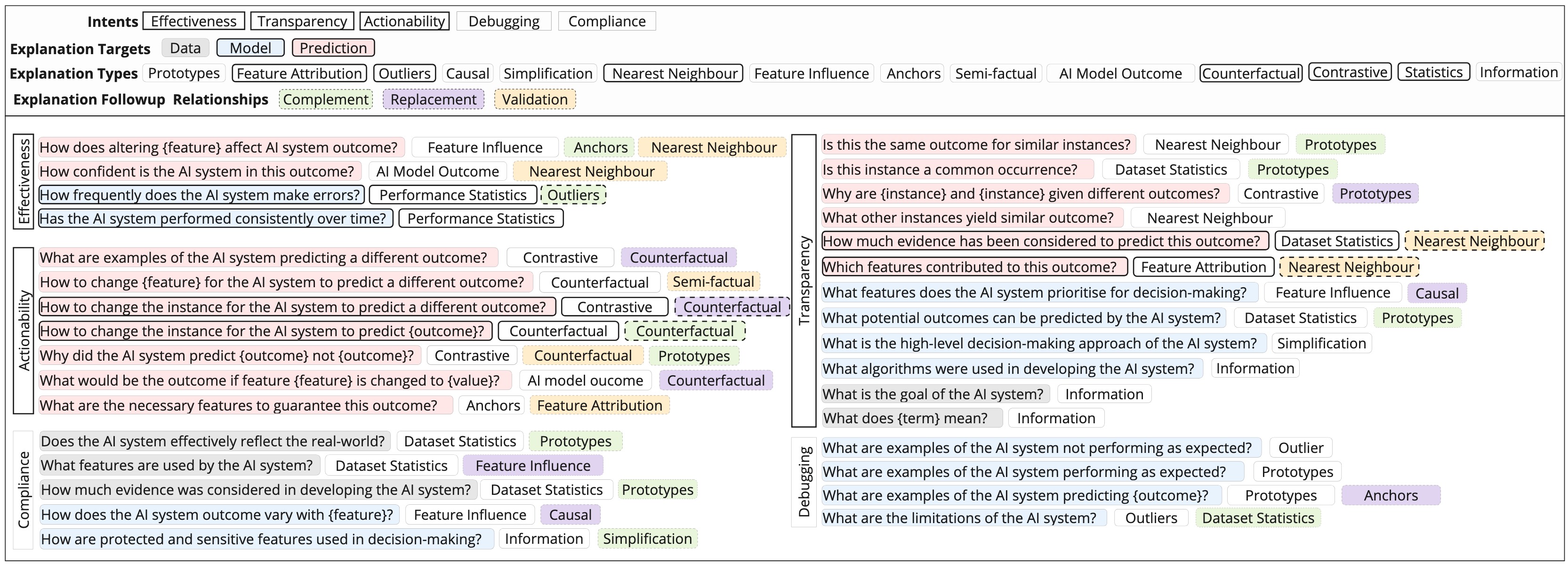}
    \caption{Intent Fulfilment Framework~(IFF). The bold black borders indicate the IFF selected for Loan Applicants interacting with a Loan Approval AI system, used as the study application in the user study.}
    \label{fig:iff}
\end{figure*}

We define an intent fulfilment ontology, providing a formal, explicit definition of the concepts and their interrelationships required to represent the necessary information, illustrated in Figure~\ref{fig:onto}. This provides a common structure and definitions, supporting wider adoption, reuse, and adaption of the IFF. 

Individuals of type \textbf{Intent} define the purpose, motivation or goal of a \textbf{User} for seeking explanations from the AI system. We define five different intents: \textit{Effectiveness}, where the user aims to understand how well the AI system performs: \textit{Actionability}, where the user aims to understand actions that need to be taken based on the AI system's output, possibly to change the output; \textit{Compliance}, where the user aims to determine the extent to which an AI system adheres to some set of principles or regulations, such as legal and ethical standards, fairness, bias mitigation, data protection; \textit{Transparency}, where the user aims to understand the decision making processes of the AI system with a focus on clarity regarding the algorithms, data sources, and overall operational mechanisms; and \textit{Debugging}, where the user aims to identify defects and limitations of the AI system.

Each intent is linked to one or more \textbf{User Question}s, that a user may ask the XAI system as part of fulfilling their intent. Each question is linked to the specific aspect (e.g. model, data, output) of an AI system that the user is seeking information about, referred to as the \textbf{User Question Target}.
\textbf{Explanation Technique}s, such as Anchors, Feature Attribution, and Counterfactual, generate \textbf{Explanation}s which aim to answer specific questions. Given the abundance of Explanation Techniques and Explanation types defined by the XAI research community, both classes are the root node in a hierarchy of more specific sub-types of techniques and explanations respectively.  

The explanation experience can be considered as the information about the user, the questions they asked, and the explanations viewed – capturing their interactions with the XAI system. This experience can be personalised to the user, not just based on their intent, but pre-existing \textbf{Knowledge} of the domain and/or AI models. 

\subsection{Intent Fulfilment Framework}

Intent Fulfilment Framework~(IFF) is comprised of multi-dimensional concepts and individuals of the ontology. We systematically curated IFF based on previous literature and input from co-design activities with industry partners. 
Figure~\ref{fig:iff} presents the IFF as a typological classification. The user questions are grouped by intent and classified considering user question targets. Each question reference one or more explanation types. These types include both well-known ones such as feature attributions from literature along with simple yet effective ones such as dataset statistics based on the input from industry partners. 
When multiple explanation types are referenced by a question, we designate one as the recommended and define Followup relations to other types. 

\subsubsection{Explanation Followup Relations}
\label{sec:followup}
We formalise the Followup relationship between two~(or more) explanation types $E_1$ and $E_2$ as either complement, replacement or validation. 

\begin{description}
\item [Complement] relationship between two explanations exists when the latter enhances or completes the understanding, scope, or functionality of the former. Together, they form a cohesive experience, each contributing unique aspects that, when combined, fulfil the intent. 
\item [Replacement] relationship between two explanations occurs when $E_2$ serves as a substitute for $E_1$, often offering contrasting information. This allows for correcting conflicts that occur when fulfilling ambiguous intents. 
\item [Validation] relationship between two explanations involves the process of confirming or verifying the accuracy, integrity, or authenticity of $E_1$ based on the information provided by $E_2$. 
\end{description}

Note that while $E_1, E_2$ can be equal, they may produce distinct explanations within an explanation experience. For instance, in the case of counterfactual explanations, $E_2$ may propose an alternative set of actionable changes to \textit{complement} $E_1$. Similarly, in scenarios involving feature attribution, $E_2$ may employ a different algorithm to $E_1$ as a means of \textit{validation}.
This explicit specification of relations between explanation types facilitates and promotes the effective reuse of the extensive array of explanation techniques developed in XAI research to date. 

\subsection{IFF Development Methodology}
The initial draft of the IFF was based on a critical review of previous literature~\cite{gilpin2018explaining,hoffman2018metrics,liao2020questioning,chromik2020taxonomy,belle2021principles,suresh2021beyond,hernandez2021convex,chen2022interpretable,chari2020explanation,arya2019one,nguyen2023black}.  
Next, we conducted several co-design workshops with two industry partners: sensor-based anomaly detection for fall detection; and upper limb radiograph fracture detection. 
Using the initial draft as the guideline they iteratively curated and refined IFFs for their AI systems. Over several co-design sessions they selected and customised user questions, assigned suitable explanation types and in the presence of multiple explanation types, identified the recommended~($E_1$) and the Followup relations of others~($E_2$) with $E_1$. Supplementary material includes anonymised screen excerpts from a Miro board created with an industry partner during their co-design sessions. We utilised the experience and outputs of the co-design workshops to further improve the IFF to present the current version. This framework is continuously improved as a general set of guidelines, we do not consider it complete or final. 

In practice, we envision a similar co-design process where domain experts and XAI researchers collaboratively curate IFFs. For instance, three XAI researchers who had experience as loan applicants curated an IFF for a Loan Approval AI system which also serves as the study application. The initial IFF selection for Loan Applicants interacting with the AI system is highlighted in Figure~\ref{fig:iff} with bold black lines.

\section{Explanation Experience Dialogue Model}
\label{sec:adf}

We extend the Explanation Dialogue Model~\cite{madumal2019grounded} to introduce an atomic dialogue type that implements Followup relationships between explanations. Following~\cite{madumal2019grounded} and~\cite{malandri2023convxai}, we formalise it using the Agent Dialogue Framework~(ADF) to introduce the Explanation Experience Dialogue Model~(EEDM).

\begin{figure}[!t]
    \centering
    \includegraphics[width=.5\textwidth]{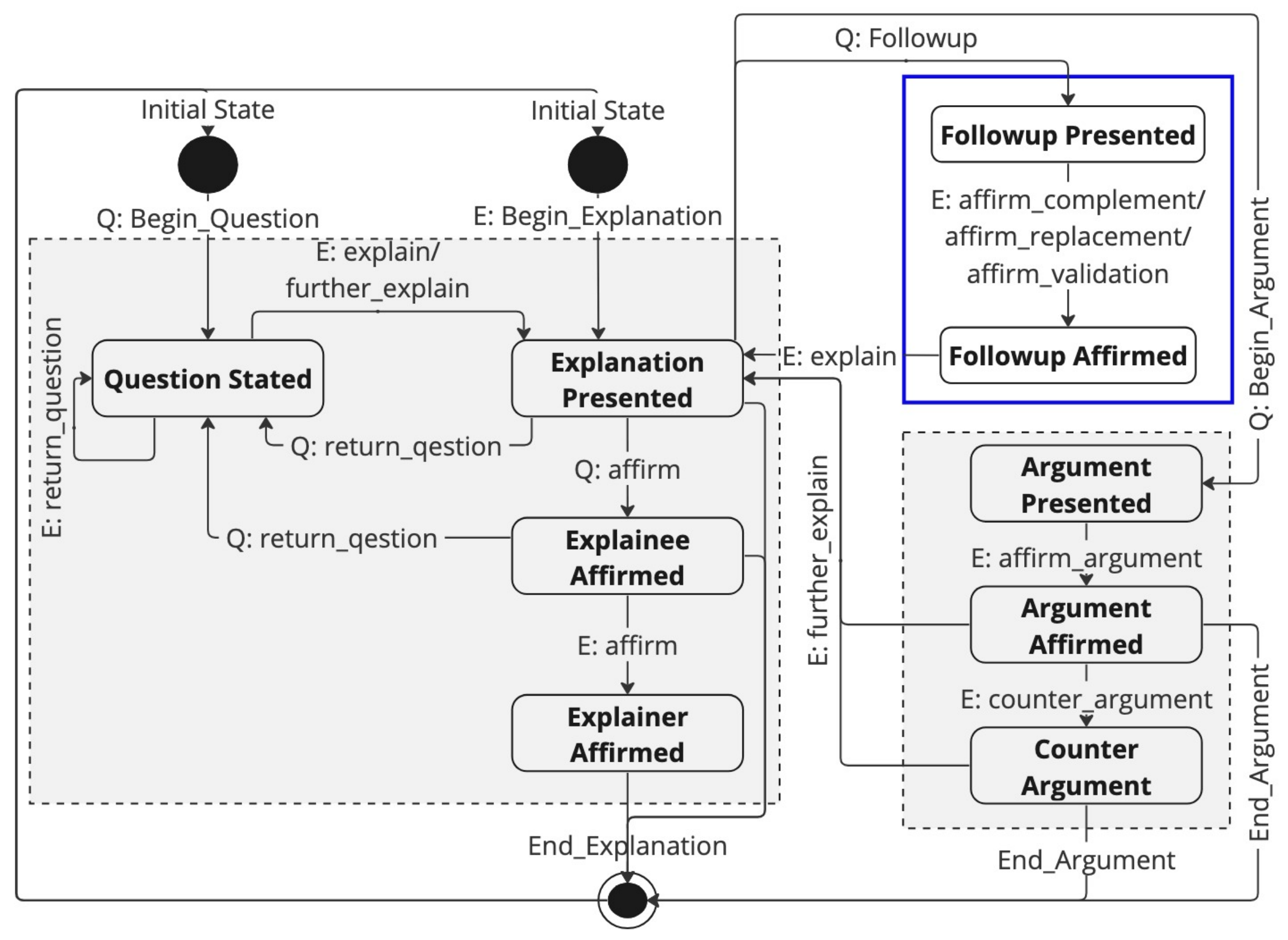}
    \caption{Explanation Experience Dialogue Model}
    \label{fig:adf}
\end{figure}

\begin{definition}[Explanation Dialogue Model]
EDM is a tuple $(\mathcal{A},\mathcal{L},\Pi_a, \Pi_c, \Pi)$ where:
\begin{itemize}
\item $\mathcal{A}$: is the set of agents $\mathcal{A} = \{\mathcal{Q}, \mathcal{E}\}$, with the labels $\mathcal{Q}$ and $\mathcal{E}$, denoting the user and the chatbot;  
\item $\mathcal{L}$ is the set of logical representations of conversation topics (denoted by p, q, r, ...);
\item $\Pi_a$ a is the set of atomic dialogue types $\Pi_a = {
G_E, G_A}$, where $G_E$ is the explanation dialogue and $G_A$ is the
argumentation dialogue;
\item $\Pi_c$ is the set of control dialogues $\Pi_c = $ \{Begin\_Question, Begin\_Explanation, Begin\_Argument, End\_Explanation, End\_Argument\}; and 
\item $\Pi$ is the closure of $\Pi_a \cup \Pi_c$ under the combination rule set. $\Pi$ provides the set of formal explanation dialogues $G$. 
\end{itemize}
\end{definition}

\begin{definition}[Explanation Experience Dialogue Model]
An EEDM is a tuple = $EEDF(\mathcal{A},\mathcal{L}',\Pi'_a, \Pi'_c, \Pi)$ that extends EDM where: 
\begin{itemize}
\item $\mathcal{L}'$ must include explanation target~($t$) and explanation intent~($i$) topics, i.e. $\mathcal{L}' = \mathcal{L} \cap \{t, i\}$; 
\item $\Pi'_a = \Pi_a \cup \{G_F\}$, where $G_F$ is the Followup dialogue type; and 
\item $\Pi'_c = \Pi_c \cup $ \{\textit{Followup}\};
\end{itemize}
\end{definition}

Next, we describe the three layers of EEDM: Topic, Dialogue and Control Layers ~\cite{mcburney2002games}.

\begin{definition}[Topic Layer]
The first layer comprises discussion topics represented by lowercase Roman letters p, q, and r. In the scope of explanation experiences, we mandate explanation target~($t$) and intent~($i$) topics. These topics influence dynamic IFF selection that initiates the dialogue as the conversation progresses. 
\end{definition}

\begin{definition}[Dialogue Layer]
The dialogue layer models atomic dialogue types in the following format: $G = \{\Theta, \mathcal{R}, \mathcal{T}, \mathcal{CF}\}$ where $\Theta$, $\mathcal{R}$, $\mathcal{T}$, and $\mathcal{CF}$ refers to permitted utterances~(i.e. Locutions), combination rules, termination rules and commitments. EEDM introduces a novel atomic dialogue type named Followup~($G_F$):
\[
G_F = \{\Theta_F, \mathcal{R}_F, \mathcal{T}_F, \mathcal{CF}_F\}\\
\]

We choose permitted Locutions~($\Theta_F$) of $G_F$ to distinguish between explanation relations. The combination rules, termination rules and commitments are illustrated in Figure~\ref{fig:adf}.

\begin{equation*}
    \begin{aligned}
    \Theta_F = \{\textit{explain}, \textit{affirm\_complement}, \textit{affirm\_replacement}, \\\textit{affirm\_validation}\}
    \end{aligned}
\end{equation*}

\end{definition}

\begin{definition}[Control Layer] Lastly the control layer specifies the state transitions leading into and out of the dialogue types depicted in Figure~\ref{fig:adf}. In addition to the transitions in EDM, we specify ``Q: Followup'' as a state transition from $G_E$ to $G_F$. ``Q: Followup'' expresses a specific type of follow-up on the recommended explanation without introducing a new intent topic~($i$), distinguishing it from ``Q: return\_question''.

\end{definition}

\begin{figure}[!t]
\centering
\includegraphics[width=.48\textwidth]{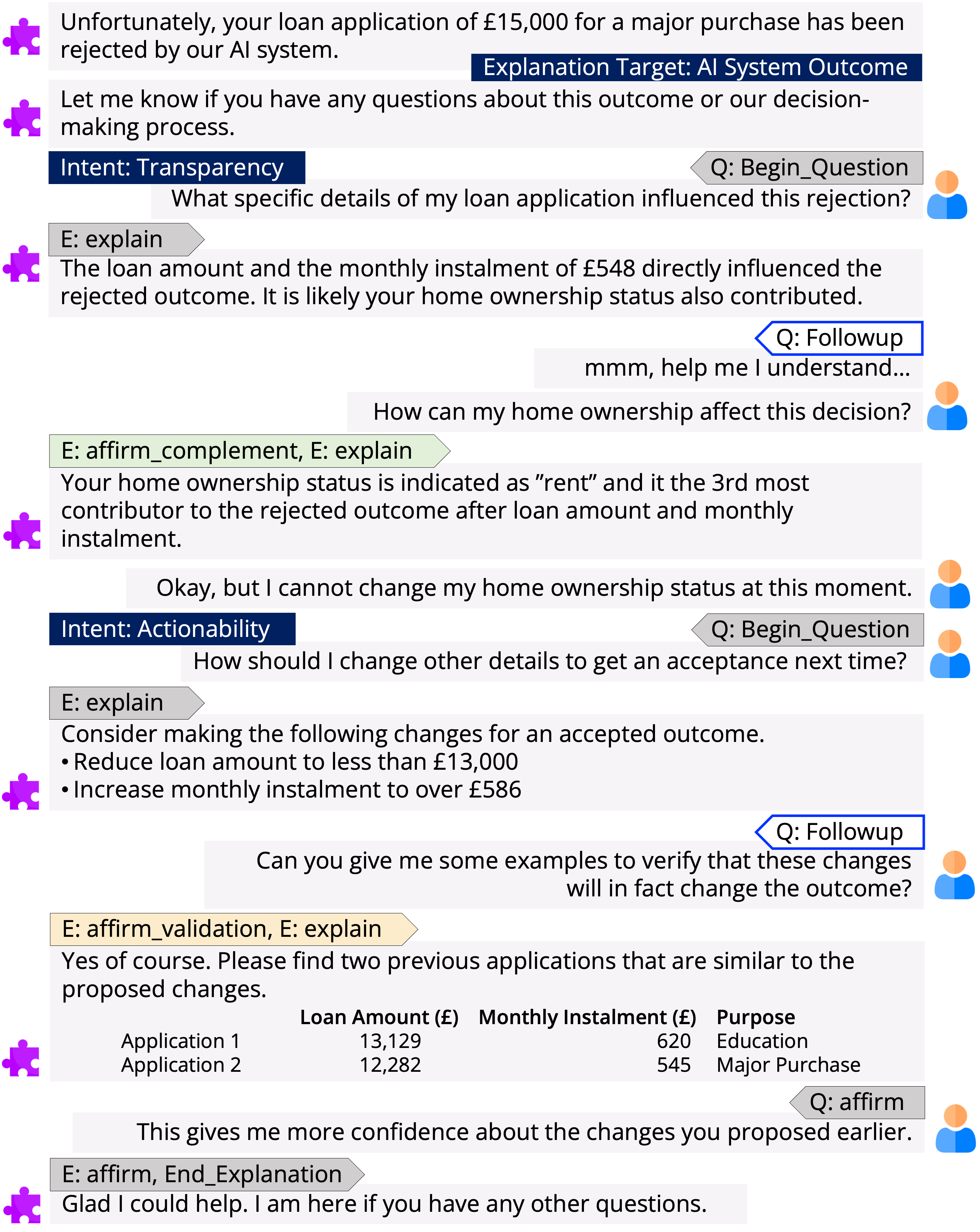}
\caption{A sample conversation between a Loan Applicant user and the Chatbot annotated with dialogue topics, locutions and references to the IFF}
\label{fig:loan}
\end{figure}

A representative sample conversation based on the IFF for the Loan Approval system use case is presented in Figure~\ref{fig:loan}. First, the chatbot presents the applicant's AI system outcome~(topic: explanation target). Thereafter the applicant goes on to explore their explanation needs. The applicant asks a question with \textit{transparency intent} and the chatbot answers with a \textit{feature attribution} type explanation. The applicant follows up by requesting \textit{complementary} information about the explanation. Their second question expresses \textit{actionability intent} answered by a \textit{counterfactual} type explanation. Before taking action, the applicant follows up to \textit{validate} the counterfactual for which the chatbot answers with a \textit{nearest neighbour} type explanation. At this stage the applicant finds their explanation intents fulfilled.

\paragraph{Implementation}
We use Behaviour Trees~\cite{colledanchise2018behavior} to implement a dialogue manager based on the EEDM. 
The atomic dialogue types are implemented as Action Nodes and the Composite and Decorator Nodes implement combination and termination rules. Dialogue Manager utilises an Explainer Library which contains at least one technique for each explanation type. In addition, NLP techniques are implemented to recognise user questions and intents and NLG techniques to generate textual annotation of images and tabular explanations. The complete Behaviour Tree and Python Dialogue Manager server GitHub are included in Supplementary Material.

\section{Evaluation}
\label{sec:study}

We conduct a comparative user study to evaluate the impact of the IFF with Explanation Followups on user experience. 

\subsection{Study Application}
\label{sec:app}

The user study is based on the well-known loan approval use case. Three XAI researchers curated the IFF for loan applicants as highlighted in bold black lines in Figure~\ref{fig:iff}. 
They selected 6 user questions across 3 intents and several explanation types. 

We configured two versions of the chatbot for comparison: \textbf{A)} when the participant asks a question, the chatbot provides the recommended explanation~(IFF without Followups); and \textbf{B)} when the participant asks a question, the chatbot provides the recommended explanation. Thereafter, the participant can choose to follow up. Considering all participants are not domain or AI experts, the complement Followup always invoked providing textual annotations for each explanation~(IFF with Followups).
Both versions allow participants to ask multiple user questions and receive explanations. Each participant was randomly assigned to a version, creating the two comparative groups A and B. 

The study starts with a set of high-level study directions. Next, participants interact with the chatbot assuming the role of loan applicants. At the end, the chatbots prompt the participants to respond to an evaluation questionnaire and provide free-text feedback. The complete conversation is automatically logged with time elapsed at each atomic interaction. 
A conversation is estimated to last for approximately 15 minutes.  

\subsection{Recruitment}
This study involved 54 participants recruited via the Prolific platform. The criteria for recruitment included: fluency in the English language; qualification of (at least) an Undergraduate degree; and holding a single/joint bank account. 90 participants attempted the study within 7 days. We rejected 17 submissions based on time spent~(16 spent $\leq$ $0.3 \times$ estimated time and 1 spent $\geq$ $4 \times$ estimated time) and 19 did not complete the study. 
Accordingly, 25 and 29 submissions were accepted for groups A and B. The mean total time spent by groups A and B is 12:36 $\pm$ 8:24 and 18:14 $\pm$ 10:54 minutes~(excluding time spent on writing free-text feedback). Participants are identified by their group prefix~(i.e. A-1, B-1) in the analysis. 

\subsection{Analysis and Results}
The analysis and findings of the study are organised into three parts: explanation satisfaction, engagement in conversation pathways and explanation experience quality.

\subsubsection{Explanation Satisfaction}
At the end of their interactions with the chatbot, participants responded to 6 questions selected from the Explanation Satisfaction Scale~\cite{hoffman2018metrics} on a 5-step Likert scale. While the scale is not designed to evaluate explanation experiences it is the most applicable quantitative evaluation metric we found in literature to establish a baseline comparison between the two groups.

\begin{figure*}[!t]
\centering
\includegraphics[width=\textwidth]{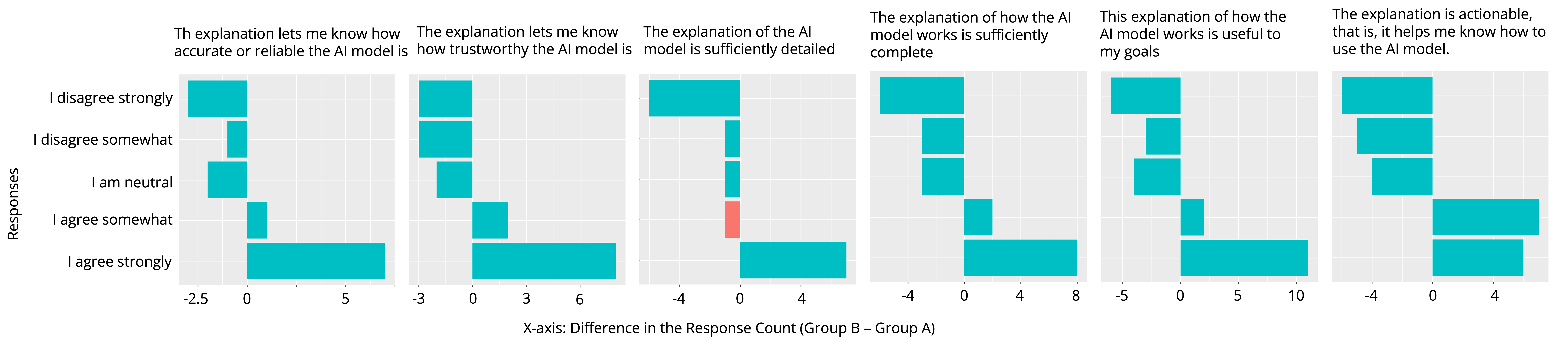}
\caption{Evaluation Questionnaire Response Difference of Group B over Group A (count B - count A)}
\label{fig:eval-responses-diff}
\end{figure*}

In Figure~\ref{fig:eval-responses-diff} we plot the difference in response counts from group A to B~(i.e. count B - count A). A Blue bar indicates a positive change, i.e. increase in positive responses and a decrease in negative/neutral responses while an Orange indicates the opposite. Overall, Group B exhibited predominantly positive responses as evidenced by the increase in positive responses and decrease in negative/neutral responses. While the exact questions and responses are specific individual interactions, we attribute the overarching positive responses from group B to the ability to follow up on the recommended explanation. Accordingly, we evidence the utility of Explanation Followups in enhancing explanation satisfaction. 

\subsubsection{Engagement in Conversation Pathways}

Secondly, we visualise the conversation pathways of participants to assess the engagement with multiple intents and specifically compare pathways in the presence (group B) and absence (group A) of Explanation Followups. 
All participants from both groups engaged with at least two intents; 2 in group A and 10 in group B interacted only with two intents. No intent was unanimously ``not chosen'' by these participants. Approximately $1\/3$ of the participants in group B did not choose a third intent which we attribute to Explanation Followups. After exploring Followups in their first two intents, participants expressed fulfilment~(further evidenced in the thematic analysis), eliminating the need to engage further.

\begin{figure}[!t]
    \centering
    \includegraphics[width=.45\textwidth]{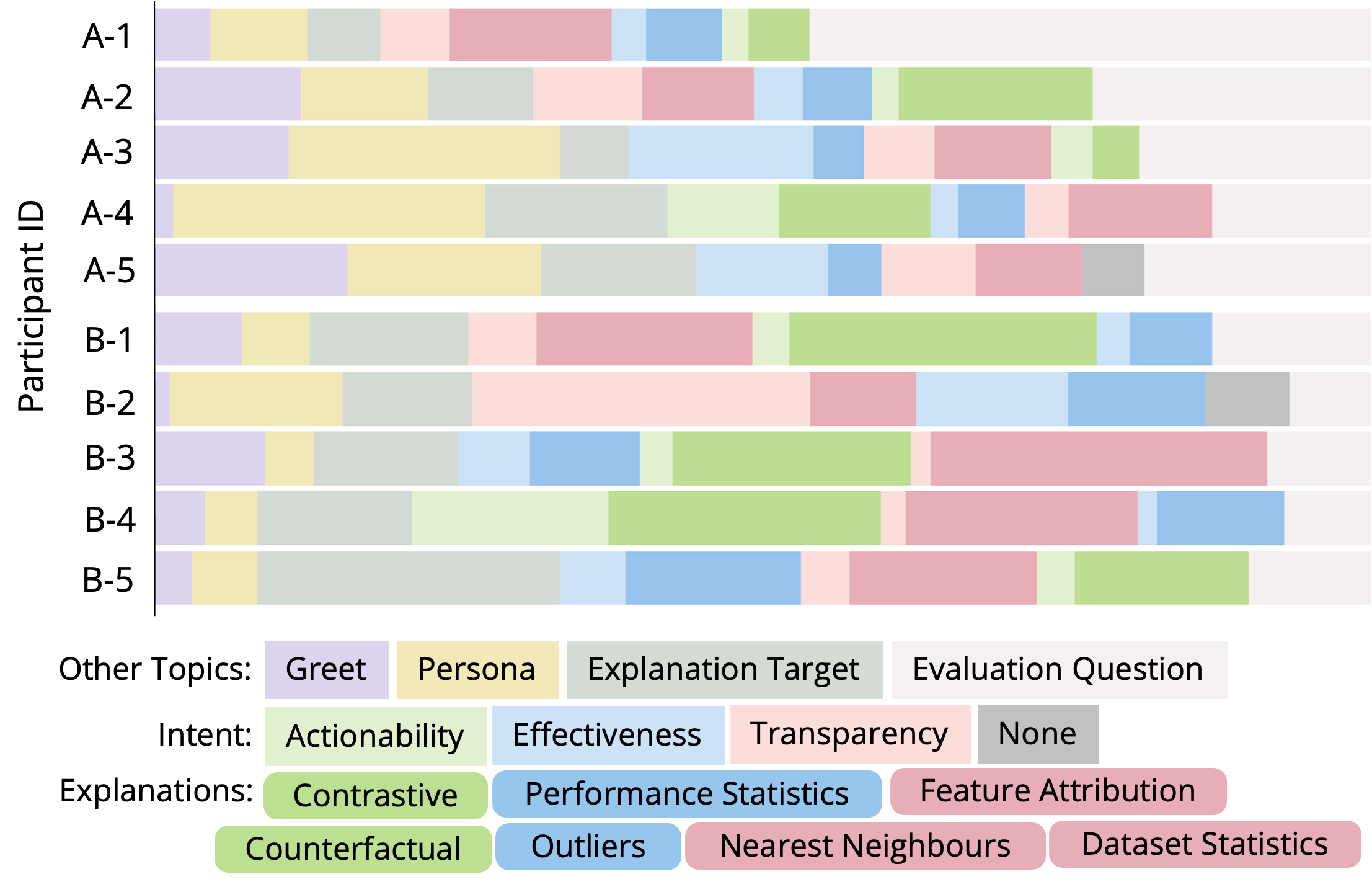}
    \caption{Conversational Pathways of participants randomly selected from Groups A and B.}
    \label{fig:iff-by-pid}
\end{figure}

For an in-depth analysis, we plot the atomic interactions of 10 participants randomly selected for groups A and B in Figure~\ref{fig:iff-by-pid}~(plots for all participants are included as Supplementary Material). The x-axis of the graph represents the sequence of atomic interactions, with each segment length representing the time spent on that interaction relative to the total interaction time. The colours of the segments correspond to dialogue topics, intents and explanation types. Participants are listed on the y-axis. To simplify the plot, we have combined the Followup interactions of group B into the recommended explanation interaction of the intent. A-5 and B-2 are examples of participants who opted out of exploring a third intent. Group B spent a higher percentage of their interaction time engaging with explanations, reflected in lengthier dark Red, Green, and Blue segments compared to Group A. This demonstrates that Explanation Followups enhance engagement within an intent while marginally diminishing the need to explore multiple intents. 

\subsubsection{Explanation Experience Quality} 
Thirdly, the free-text feedback provides qualitative evidence on the impact of the IFF and Explanation Followups.
Participants responded to \textit{``Please describe your explanation experience (min. 100 words).''} and where the mean response length was $122.89\pm 37.65$ words. We conducted a deductive thematic analysis~\cite{fereday2006demonstrating,pearse2019illustration} to compare the experiences between the groups. 

Explanation experience quality dimensions set the theoretical framework of the deductive thematic analysis. The 3 dimensions \textit{Learning}, \textit{Utility} and \textit{Engagement} were compiled and validated by a group of XAI experts who also provided the initial codebook for each dimension. Two researchers annotated the responses using the initial codebook. Next, they reviewed the annotations to develop a shared codebook with supporting annotations representing both positive and negative sentiments. We finalised 3 themes~(aligned with the three dimensions), each connected to $5-7$ codes and supported by quote excerpts from the two groups. 

\paragraph{Learning} was defined as the extent to which the experience developed knowledge or competence of the AI system. It was described using codes such as comprehension, understanding and learning. 
Group A expressed that the explanations were too technical and that it impacted their understanding. For instance, A-6 said \textit{``[explanation]it remains rather technical and it is likely that the majority would not understand the specifics''}. Group A shared the sentiment that the lack of personalisation of explanations to the audience impacted their understanding: \textit{``interesting but maybe assumes too much knowledge from the user''} (A-16). 
A-19, A-21 and A-11 specifically called for textual annotations as a means to improve learning: \textit{``there could be more written feedback to provide the user with more comprehensible detail on how the AI is shaping it's outcomes''} (A-21). 

Group B shared the sentiment that complementary information and other Followups contributed to an enhanced learning experience~(B-8,B-10,B-16,B-29).
\begin{quote}
\textit{``answers were not only detailed in the words, but also used graphs, figures and images to explain the statistics behind the reasoning. I did not expect that level of detail although it was needed because of my lack of knowledge...''} (B-19)
\end{quote} 
The availability of multiple explanation types on in IFF was positively received: \textit{``use of different explanation techniques, such as Counterfactual and Kernel SHAP, provides a comprehensive view of how AI models arrive at decisions. ''} (B-23).
Several participants reiterated or critiqued the explanations they received, indicating in-depth understanding. B-20 reiterated their interpretation of the explanations as: \textit{``LIME visualizes how qualities influence decisions, whereas Counterfactual depicts attribute alterations that potentially change outcomes. The employment of these strategies improve transparency.''}. B-3 reflected on their experience and critiqued the explanations: \textit{``I was also surprised that the reason for the loan didn't have a larger attribute, especially a 'major purchase' is a much safer bet, compared to a holiday in which case I would say much less safe bet.''}.
Overall, there was a significant difference in learning experiences between groups, with the positive impact of the Explanation Followups evident throughout the comments from Group B.

\paragraph{Utility} refers to the contribution of the experience towards task completion and/or achievement of user goals. This theme is supported by codes such as accuracy, usefulness, trust and confidence.
Group A indicated a lack of validation and verification as factors that diminished the utility. A-15 expressed the need to validate the explanations in order to establish trust: \textit{``as to someone unfamiliar with the topic like me, I can't guarantee that what has been sent is accurate''}.
A-22 expressed the need for more information to validate the AI outcome before taking action: \textit{``I'd like it to give some more information about why I was rejected''}. A-16 could envision utility as \textit{``could be helpful to me in the future''} but only if the content is tailored to their knowledge level. 

Group B expressed confidence and trust to use the tool in future and recommended who may benefit from such a tool. 
\begin{quote}
\textit{``from a transparency perspective that the level of detail provided is so high as it will give people an unbiased understanding of why they have been accepted or rejected and actions they can take''} (B-26)
\end{quote}
A similar sentiment was shared by B-22 saying that \textit{``it empowers loan applicants to comprehend the rationale behind the decisions''}.
B-15 and B-29 envisioned the utility of the tool for loan applicants. B-29 found it useful for self-assessment before applying and B-15 found the counterfactuals as a way to find suitable modifications: \textit{``the ability to make modifications to your loan application and see how that would affect the decision-making process would likely come in very handy''}. B-2 envisioned the tool being utilised in the future within the industry, stating: \textit{``would help any banking institution or lenders in providing a clear method for approving loans - aid customer service and satisfaction''}. Compared to Group A, Group B recognised the utility of the AI system as a result of improved explanation experience in task completion and achieving user goals.

\paragraph{Engagement} refers to the quality of the interaction between the user and the system and the supporting codes include experience, engagement and accessibility. 
Both groups agreed that IFF with multiple intents and user questions contributed to creating positive experiences. 
\begin{quote}
\textit{``the way the workings are explained through numerous questions that you can independently ask is a good move''} (A-20)
\end{quote}
A-18, A-19 and A-4 called for improved accessibility, specifically advocating for additional textual explanations. Both expressed that accessibility will improve the engagement to a wider audience:  \textit{``using less technical language would make the AI more accessible to a wider audience''} (A-4). 
Furthermore, A-19 reiterated how textual explanations contribute to more dialogue-like interaction. 

Group B reported improved quality of experiences directly attributing to the IFF and specifically Explanation Followups. 
\begin{quote}
\textit{``dynamic approach ensures users remain engaged and gain a comprehensive understanding of the system's workings''} (B-22)
\end{quote}
Many shared the sentiment that the pre-configured IFF mitigates cold-start and encourages continued engagement: \textit{``suggested questions were a nice touch, especially with deciding which steps I could take next''} (B-23).
Overall Group B's sentiment towards engagement can be summarised by a quote from B-25: \textit{``...chatbot offers an intriguing user-friendly model which keeps the client active throughout the process''}.

\subsubsection{Implications and Limitations}

The impact of the IFF design and Explanation Followups were evident through our evaluations. Group B, who interacted with Followup dialogue showed significant improvements in satisfaction, engagement, and overall experience quality. Users with limited domain knowledge benefited from complementary explanations, leading to an enhanced learning experience, while validations increased the AI system's utility. Consequently, both qualitative and quantitative evidence demonstrated improvements in the overall experience quality.

While the above conclusions are generalisable, we acknowledge a key limitation that warrants improvement through future work. 
The evaluation will expand to encompass multiple use cases and IFF configurations, catering to diverse user groups with varying levels of domain expertise.
We envision this broader scope will enable 1) identifying patterns and preferences in Explanation Followups among diverse user groups towards developing an ontology-based approach to formalising Followup relations; and 2) incorporating domain knowledge levels as a typology in the IFF. Both these improvements will enhance the utility of the IFF as a set of guidelines for developing user-centred XAI systems that comply with explainability regulations and promote good practices. 

\section{Conclusion}
\label{sec:conc}
This paper presented the Intent Fulfilment Framework~(IFF) that integrated the works of previous research in Multi-faceted XAI and insights from the industry for creating explanation experiences.  
Through the introduction of Explanation Followup relations, now users can seamlessly delve deeper into explanations, seeking additional information, alternative explanations and/or validation.  This explicit delineation of \textit{Followup} types not only promotes the reuse of explanation techniques from literature but also reduces reliance on expert knowledge. 
We formalised the Explanation Experience Dialogue Model~(EEDM) that models and implements IFF and Explanation Followups into conversational explanation experiences.
Our evaluations demonstrated significant improvements in the quality of explanation experiences through the IFF and Explanation Followups.
They experienced enhanced learning experiences, with reinforced trust and confidence, leading to increased utility of the AI system and improved engagement.

\appendix

\section*{Ethics Approval}
Both co-design activity and the user study protocols passed the ethics review of the leading institution~(references removed for review). Informed consent was obtained from all participants.


\appendix

\section{Dialogue Manager Implementation Details}
\begin{figure}[ht]
    \centering
    \includegraphics[width=.5\textwidth]{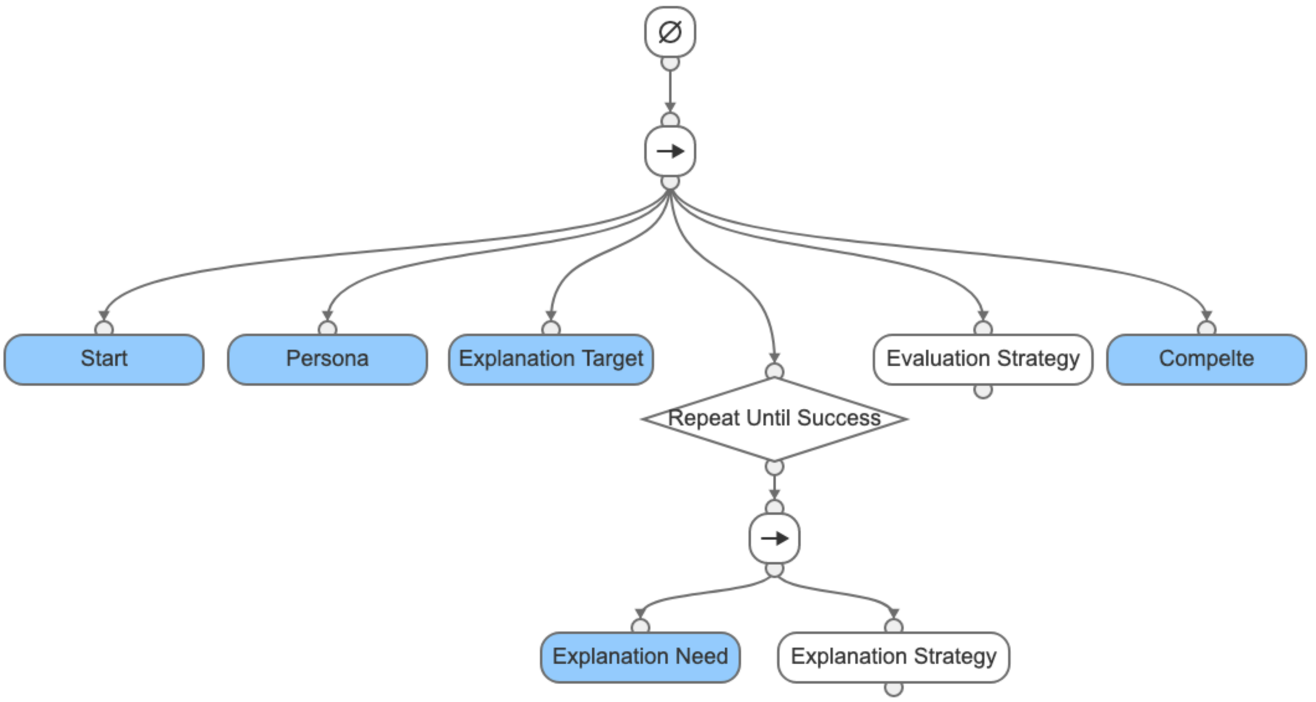}
    \caption{Abstract BT of the Dialogue Manager}
    \label{fig:bt}
\end{figure}

We transform the Explanation Experience Dialogue Model into a Behaviour Tree~(BT) from which we implement a dialogue manager. Figure~\ref{fig:bt} presents the abstract BT of the dialogue manager. 

\begin{description}

\item[Persona Node] interacts with the user to establish their knowledge levels and identifies the user group they belong to within the use case context. 

\item [Explanation Target Node] interacts with the user regarding their data and the AI decision~(e.g. the data instance and the AI prediction). The node can be configured to upload a specific type of data instance, obtain feature values in a question-answer manner or randomly select from test samples provided by the design users. 

\item [Explanation Need Node] implements an interaction where the user explores their explanation needs after receiving the AI decision. Explanation needs~(i.e. intents) are expressed as questions. 

\item[Explanation Strategy Node] is a placeholder composite node to plug and play explanation strategies based on the use case, user group and explanation target and intent. An explanation strategy is designed using explainer nodes, condition nodes and Followup composite nodes that handle navigation from left to right sibling explainer nodes.

\item[Evaluation Strategy Node] is a placeholder composite node for evaluation metrics~(i.e.\ questionnaire). It maintains a sequence sub-tree where new Evaluation Question Nodes are appended as new explanation needs are encountered during a conversation.
\end{description}

\subsection{Dialogue Manager Implementation Architecture}
\begin{figure}[ht]
    \centering
    \includegraphics[width=.5\textwidth]{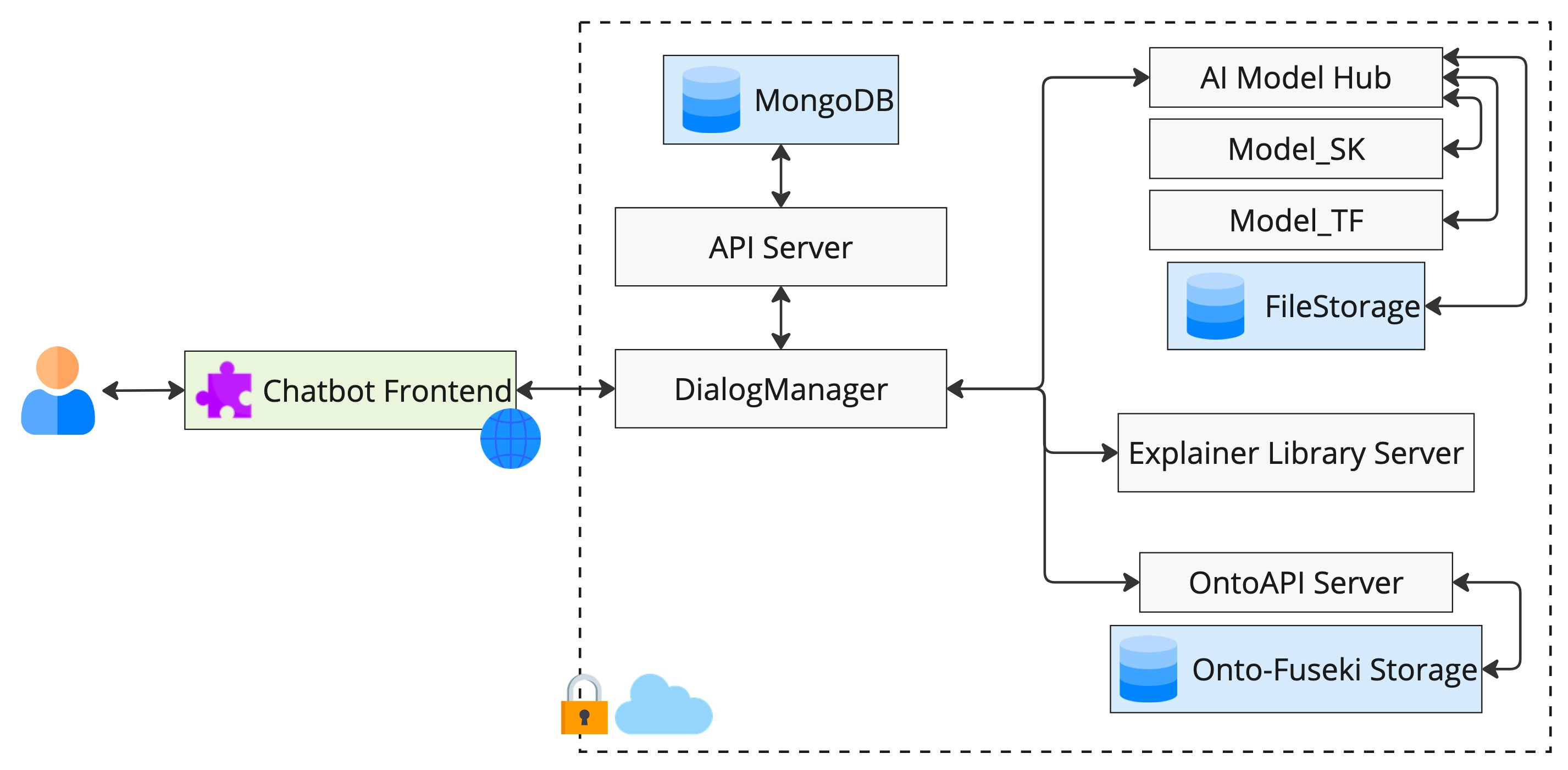}
    \caption{Dialogue Manager Implementation Architecture}
    \label{fig:arch}
\end{figure}
The dialogue manager manages the conversation by executing the BT. When a user initiates a conversation, the dialogue manager maintains the connection with the front end through a secure WebSocket.
AI model hub provides access AI models, and data and generates predictions. The Explainer Library is comprised of a collection of explanation techniques and generates explanations. 
Access to the IF ontology and the IFF stored in the Fuseki triplestore is facilitated through the OntoAPI. The API server and MongoDB manage authentication and are utilised to store the conversation logs. The complete system is dockerized and hosted on a cloud environment.

\section{IFF Co-design Activities}
\begin{figure*}[!ht]
    \centering
    \includegraphics[width=.9\textwidth]{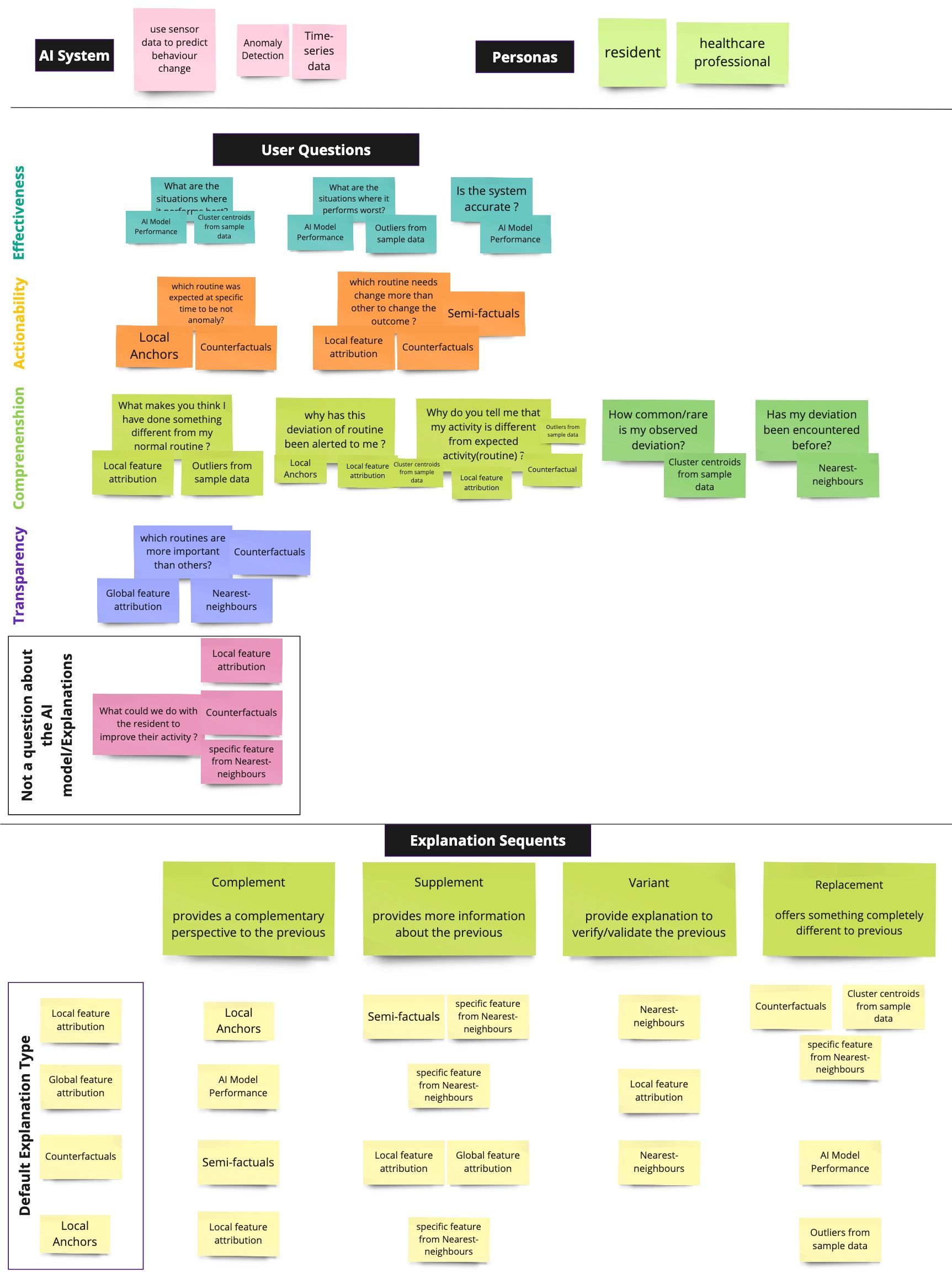}
    \caption{Output from co-design activities with the industry partner use case on sensor-based fall detection. The content has been anonymised, colours updated and organised to fit the page and for readability}
    \label{fig:miro}
\end{figure*}

Figure~\ref{fig:miro} presents the screen excerpts from the Miro board created with the industry partner who was looking to implement explainability for their sensor-based fall detection AI system deployed in smart homes for assisted living. We refer to the industry partner as the design user. Following are the co-design tasks we carried out to produce this Miro board. 
\begin{description}
\item [Design User] Describe the AI model. Anomaly detection AI model implemented using Neural Network trained on time series data labelled for ``fall'' or ``nominal activity''. 
\item [Design User] Describe user groups. The two user groups interested in explaining the AI model or the predictions it produces are the residents of the home or healthcare professionals treating/monitoring residents. Only healthcare professionals have domain knowledge, not the residents.
\item [Design User and Researcher] Explore the given IFF to select and customise the questions the two user groups may have about the AI system or the predictions. 
\item [Design User and Researcher] Annotate what intents the questions belong to and what explanation types can answer the questions. 
\item [Design User and Researcher] When multiple explanation types answer one question, one is selected as recommended and specifies the relationship of other explanation types with the recommended. 
\end{description}

Note that at the time of this co-design workshops, we considered 4 Explanation Followup relations which were later revised to 3 combining \textit{Complement} and \textit{Supplement}. Similarly, we considered Comprehension and Transparency to be two distinct intents instead of one in the current version of the IFF.

\section{User Study: Extended Material and Evaluations}
\begin{figure*}[ht]
\centering
\includegraphics[width=.6\textwidth]{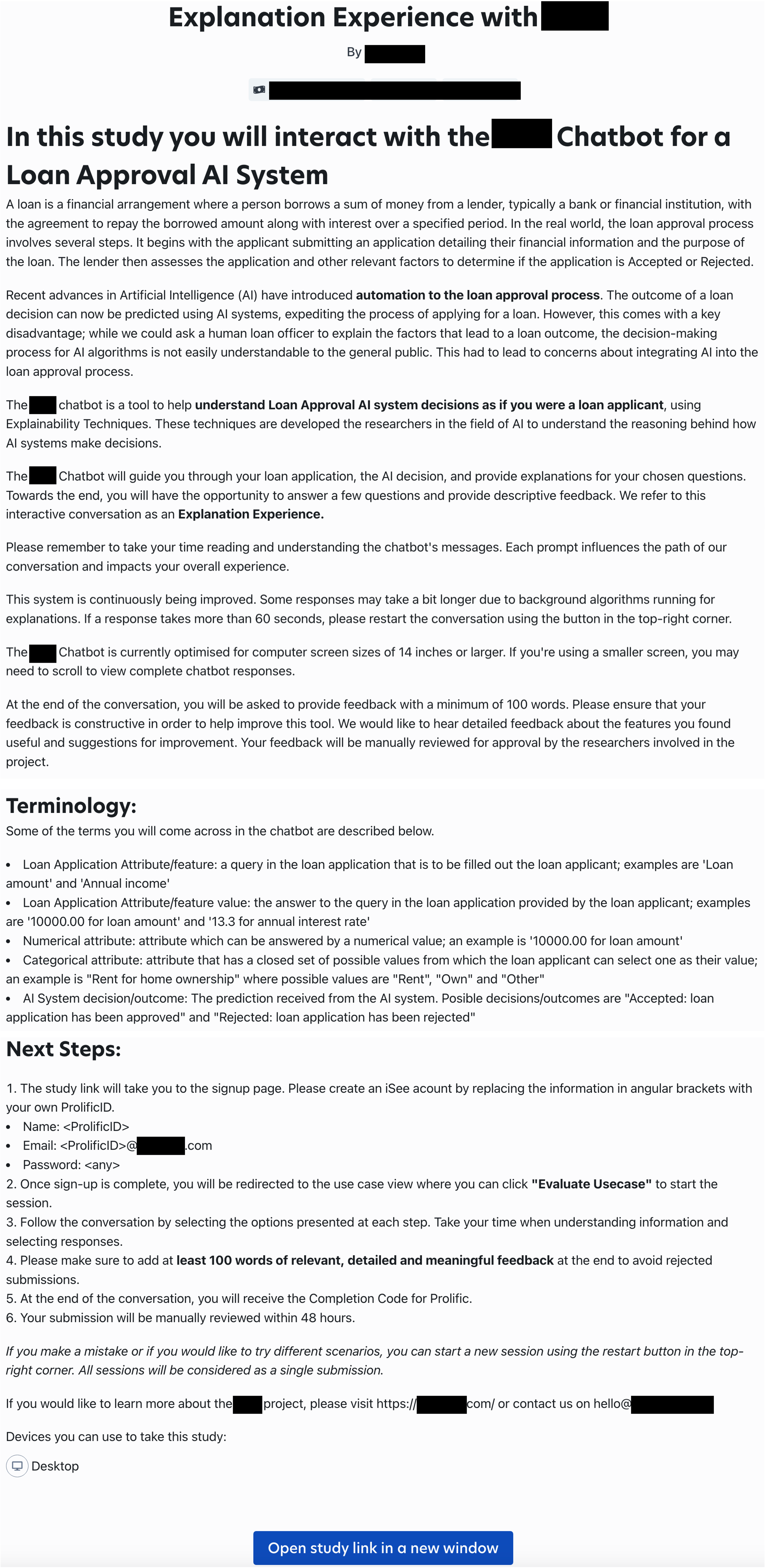}
\caption{Study instructions on Prolific.co platform}
\label{fig:prolific}
\end{figure*}

Figure~\ref{fig:prolific} presents a screenshot of the study instructions provided before the user study on Prolific.co platform. Some information is redacted for anonymity. The participants were provided with an introduction to the application domain and a description of what to expect when they started the study before they proceeded to the chatbot. 
\subsection{Study instructions on Prolific}

\subsection{Example Conversation}
\begin{figure*}[!hbt]
    \centering
    \includegraphics[width=.8\textwidth]{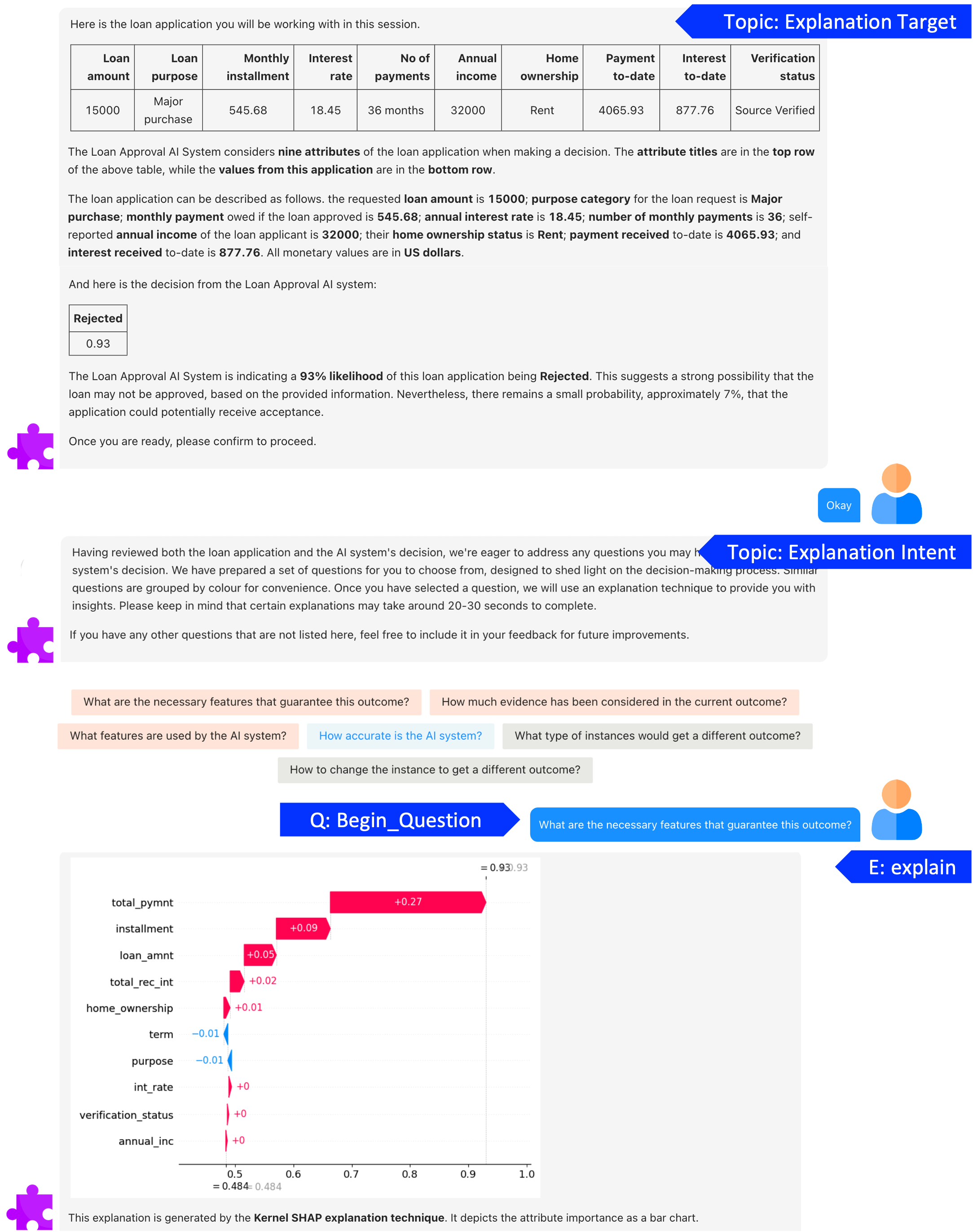}
    \caption{Sample of Loan Applicant's explanation experience. The sample continues to the next page...}
    \label{fig:loan1}
\end{figure*}

\begin{figure*}[!ht]
    \centering
    \includegraphics[width=.8\textwidth]{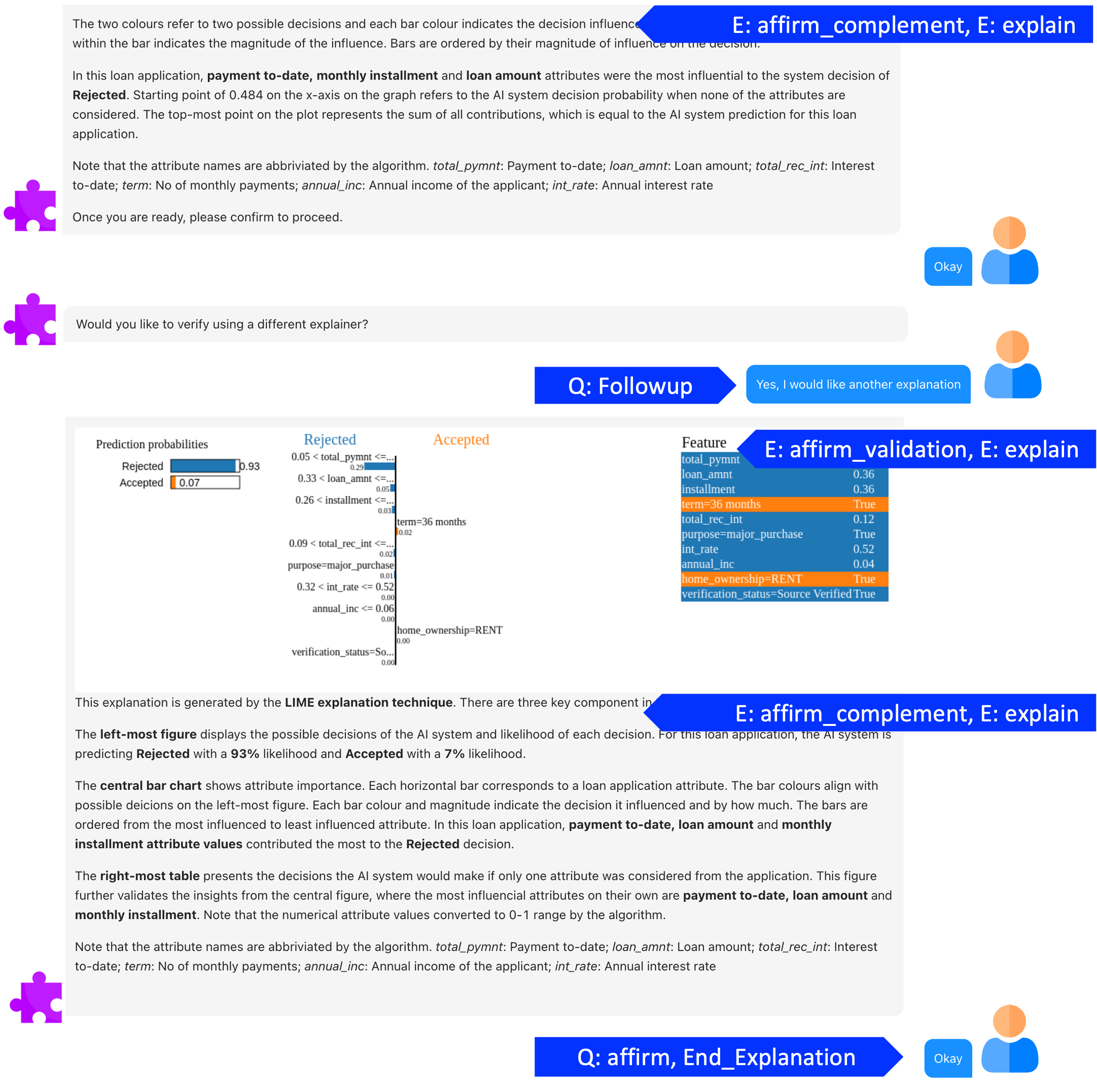}
    \caption{Sample of Loan Applicant's explanation experience continued from the previous page. }
    \label{fig:loan2}
\end{figure*}

Figure~\ref{fig:loan1} and~\ref{fig:loan2} presents a sample conversation from Group B. These figures have been reconstructed from conversation logs and replaced the profile pictures for anonymity. In comparison, Group A only received the recommended explanation as it was generated from the explanation technique implementation and did not receive any complementary textual annotations or opportunities to follow up. 

\subsection{Explanation Satisfaction: questionnaire responses}
\begin{figure*}[!ht]
\centering
\includegraphics[width=.9\textwidth]{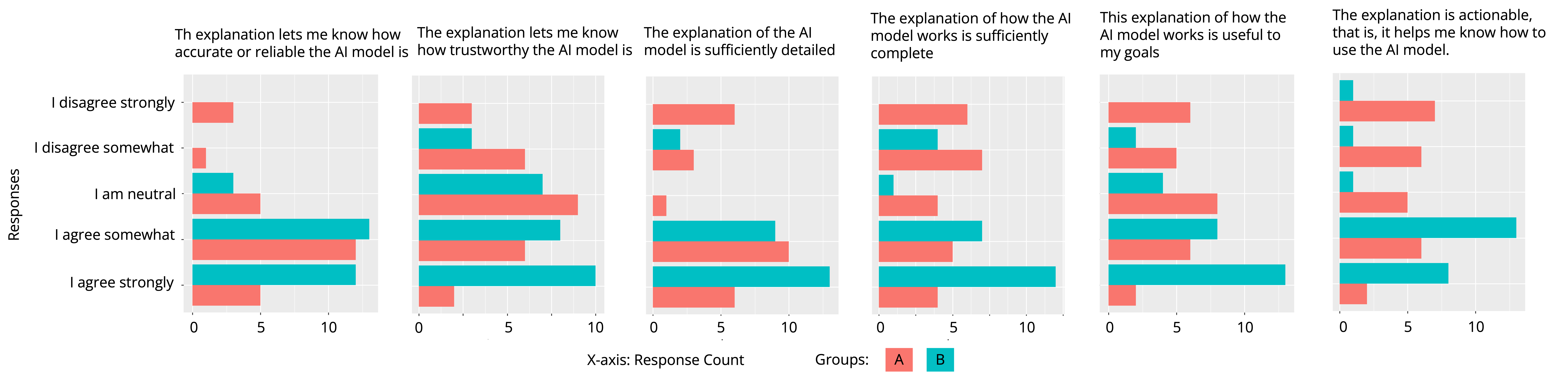}
\caption{Evaluation questionnaire response counts compared between groups A and B}
\label{fig:eval-responses}
\end{figure*}

Figure~\ref{fig:eval-responses} presents the response counts received from groups A and B on the evaluation questionnaire. 

\subsection{User Engagement: Atomic interactions of all participants}
\begin{figure*}[!ht]
\centering
\includegraphics[width=.9\textwidth]{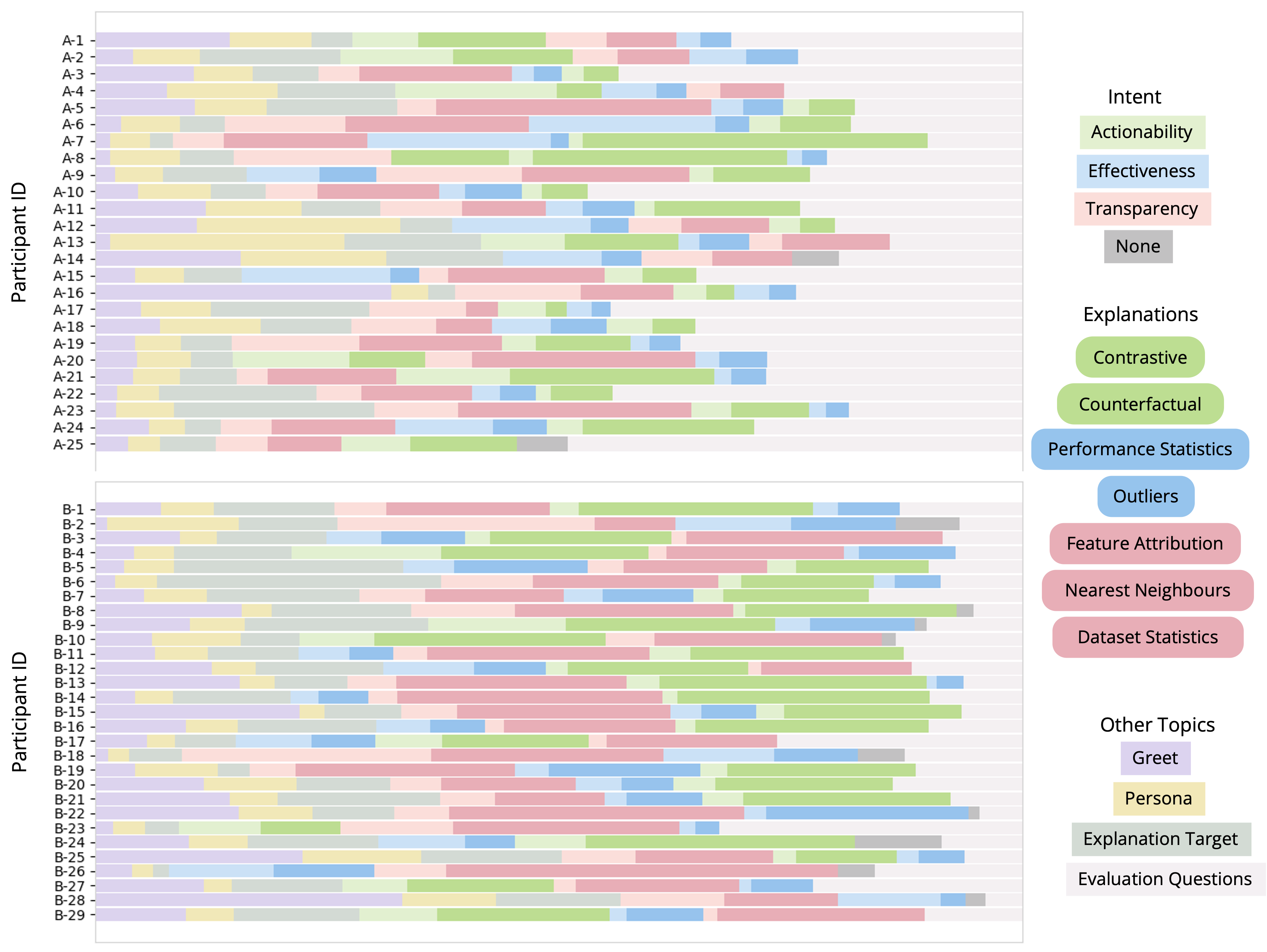}
\caption{Conversation pathways of Groups A and B: Time elapsed at each atomic dialogue as a percentage of the complete conversation}
\label{fig:iff-per}
\end{figure*}

\begin{figure*}[!ht]
\centering
\includegraphics[width=.9\textwidth]{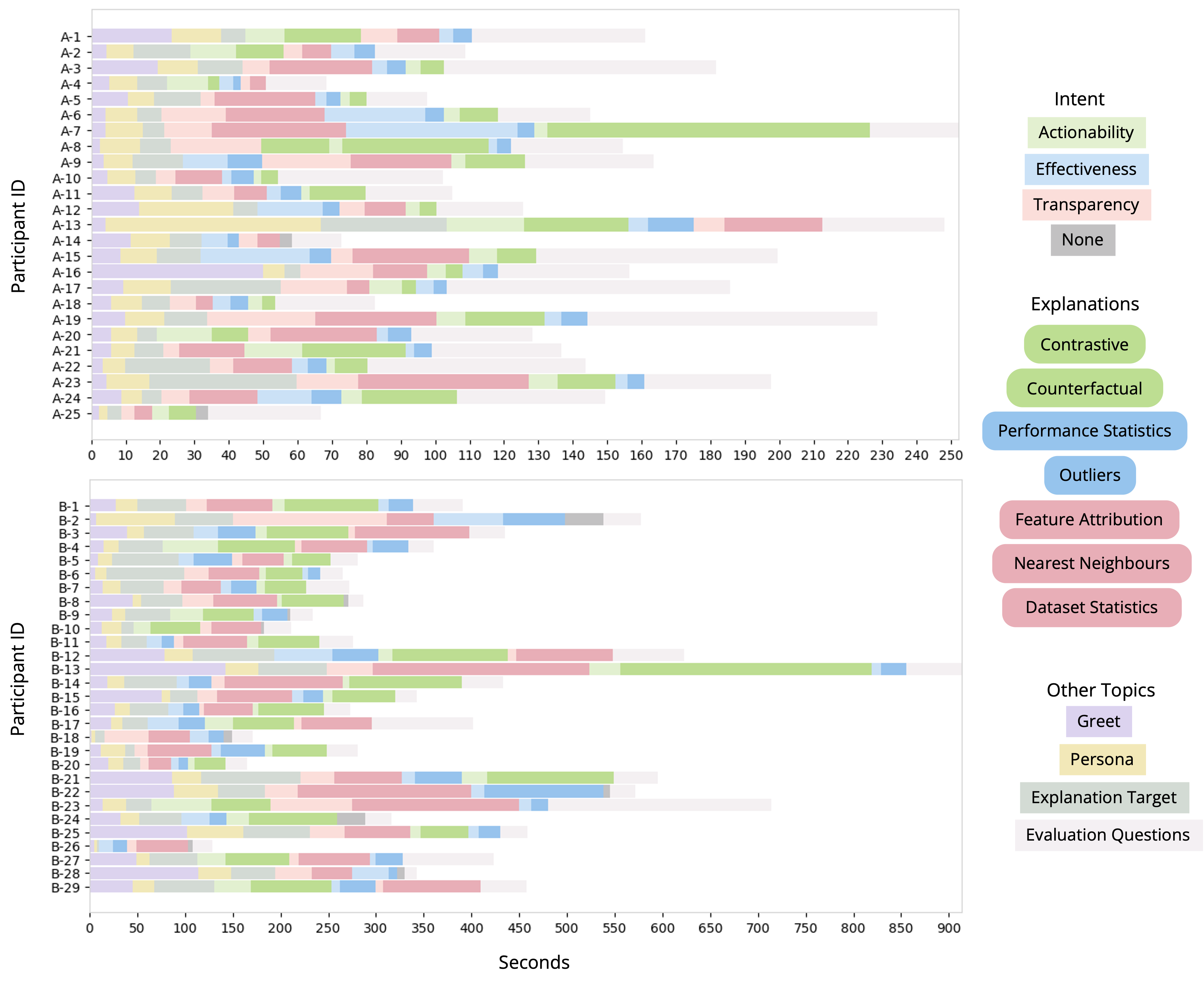}
\caption{Conversation pathways of Groups A and B: Real-time elapsed at each atomic dialogue in Seconds}
\label{fig:iff-real}
\end{figure*}

Figure~\ref{fig:iff-per} presents the complete conversation pathways of the two groups where the segment length is relative to the time they spent on the complete conversation.  
Figure~\ref{fig:iff-real} presents the same conversation pathways of the two groups but with the real-time spent on atomic interactions.  These interaction logs help us to identify participants who paused and resumed the study such as A-7 and B-13.  
\label{ap:evale}

\clearpage
\bibliographystyle{named}
\bibliography{ijcai24}

\end{document}